# Study of a Passive Gating Grid for Ion Back Flow Suppression

**V Zakharov, P Garg, T Hemmick, K Dehmelt**
Department of Physics and Astronomy, Stony Brook University, Stony Brook 11794-3800, USA

E-mail: vladislav.zakharov@stonybrook.edu

**Abstract**. Space Charge (SC) distortions are some of the main issues for high-resolution Time Projection Chambers (TPC). The two main SC sources are those from primary ionizations and those that result from amplification stages. The gain stages are required to increase the electron ($e^-$) signal above electronics noise levels, but this inevitably creates extra ions. These ions can enter the drift region and distort the electric field, and thus lower the detector performance. We will present a brief motivation for our Ion Back Flow (IBF) studies along with explanations of existing techniques and our simulation results to reduce IBF. We propose several mesh structures along with static bi-polar gating. Further, we discuss position distortions in $e^-$ trajectories due to a static bi-polar grid and use these distortions to compensate for non-linear responses of our Zig-Zag pad readout.

## 1. Introduction

The upgraded sPHENIX experiment for the Relativistic Heavy Ion Collider (RHIC) in Brookhaven National Lab (BNL) is designed to study jet measurements, b-quark tagging, and to resolve Upsilon 1s, 2s, 3s states. A TPC is one of its main tracking detectors, which measures space points of charged tracks. It will provide position resolution below 200μm to separate the Upsilon states through its dielectron decay channel. This will help to study the time evolution of the Quark Gluon Plasma (QGP). Our studies may also be useful for IBF reduction in TPCs to be used in future collider experiments.

The sPHENIX TPC covers a full azimuthal ($\varphi$) range and pseudo-rapidity ($\eta$) of |1.1|. The total length is about 2m and the radius spans 20-to-78cm. A central membrane with flexible copper-stripped Printed Circuit Board (PCB) cards create a uniform drift field of 400 V/cm. The TPC is enclosed by a superconducting solenoid magnet creating a magnetic field of 1.4T. The read-out plane has specialized detection pads with a uniquely overlapping Zig-Zag shape of 2mm pitch and 12.5mm length along the radius, and they are specifically designed for high position resolution by maximizing charge sharing. Numerous studies have been performed on their properties and capabilities, and sPHENIX will be the first detector to benefit from advancements in eliminating their Differential Non-Linearity (DNL), which is a measure of deviation from the expected result across the Zig-Zag pattern.

As a demonstration of the IBF problem: for a gain stage of 2,000 and IBF of just 1%, the amplification leads to 20 ions coming from the gain stage compared to only 1 ion from the primary $e^-$. From this, IBF accounts for 95% of the total SC. Both ion and $e^-$ transport in electric ($\vec{E}$) and magnetic ($\vec{B}$) fields are described by the Langevin equation of motion, as follows:

$$m\frac{d\vec{v}}{dt} = q\vec{E} + q(\vec{v} \times \vec{B}) - \kappa\vec{v} \quad (1)$$

Here m, q, $\vec{v}$ are the mass, electric charge, and velocity of the particle respectively. The kappa $\kappa$ term characterizes the drag, or friction, which is the longitudinal part of the diffusion. Since ions are much heavier than $e^-$, and are oppositely charged, they will have a much smaller velocity and will move in the opposite direction as compared to the $e^-$. Their low velocity means that they will not feel an appreciable magnetic force, as compared to the fast $e^-$.

We begin by examining the effect from just the $\vec{E}$–fields on the charge transport. There are several gain structures where the $E$-drift, which is the region prior to the gain stage, is lower than the $E$-transfer, which is the region after the gain stage. Since $\vec{E}$-field lines can never cross and must start or end on charges, most of the drift lines, which the $e^-$ follow, will make it through the gain region. The ions will follow the transfer lines and a fraction of them will terminate on the top surface of the gain stage. This has been studied in simplified simulations without diffusion [5], such as in Gas Electron Multipliers (GEM), shown in Fig. 1a. Similarly, based on this principle of raising the field ratio to lower the IBF, data has been collected for a Micromegas (µMega) detector [3], and its results can be seen in Fig. 1b.

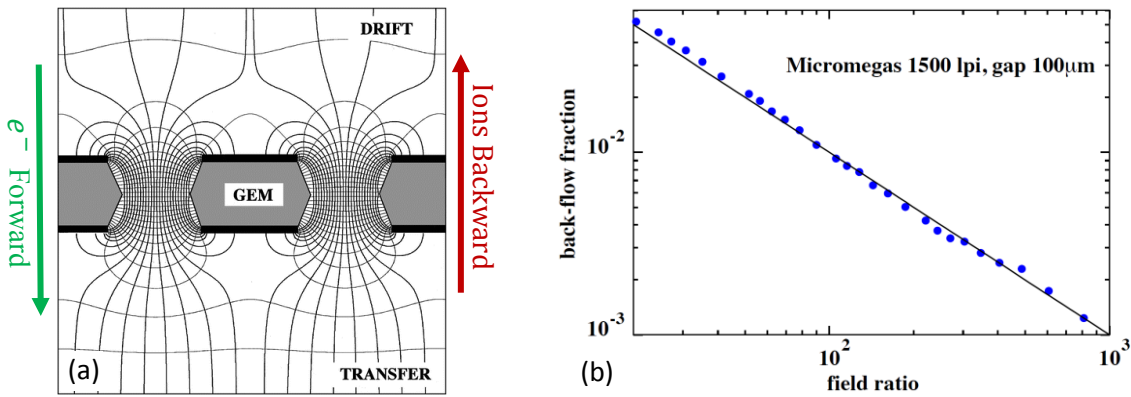

**Figures 1**: (a) Electrostatic GEM simulations [5],
(b) µMega detector data showing the IBF fraction as a function of field ratios [3]

GEMs and µMegas have frequently been used as gain structures to decrease IBF [2, 4, 7, & 8]. The applied voltages to different stages of their gain can control the field ratios and provide quite low IBF. However, this can cause other undesirable affects, such as resolution loss. As alternatives, we will now discuss passive wire & etched mesh simulations. Then, in section 3, we discuss a bi-polar static wire grid. In section 4 we discuss a modified pad shape, and in the last section we conclude our findings.

## 2. Passive Wire & Etched Mesh Simulations

The principle of a large transfer to drift field ratio to provide low IBF applies even without gain, and this allows for the use of passive structures. We study the effects of $\vec{E}$ and $\vec{B}$ fields on Electron Transparency, which is the percentage of primary $e^-$ that pass through the passive structure to the gain stage, along with Ion Blocking, which is the percentage of the ions from the gain stage that are prevented from entering the drift region, by the passive structure. Several passive mesh configurations are examined, which are simplified models of woven meshes. Fig. 2 shows two of our models that produced the best results.

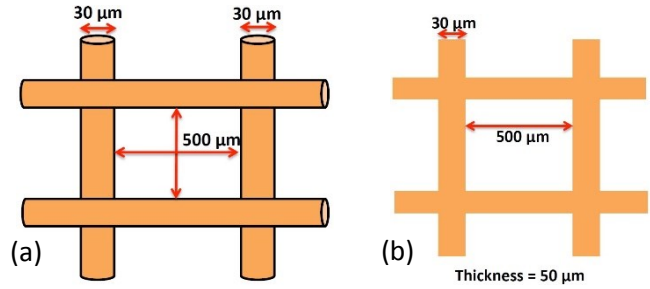

**Figure 2**: (a) Simple Wire Mesh & (b) Etched Mesh configurations

We used this model to get the electric field map for various settings using ANSYS finite element simulation package [9]. These $\vec{E}$-field maps are transported into the CERN based Garfield++ simulation set-up [10]. Garfield++ facilitates to implement the $\vec{B}$-field, gas properties, and charge transport. We explore different dimensions and $\vec{E}$-field settings to optimize the $e^-$ transparency and ion blocking through these structures.

We've scanned transfer and magnetic fields to examine the structures in Fig. 2 and one of the best performing results is shown in Fig. 3. The gas used is Ne:CF$_4$:iC$_4$H$_{10}$ (95:3:2) with $\vec{E}_{drift}$ = 400V/cm. The optical transparency ~90%. At field ratio of only 3 almost all the $e^-$ are able to pass, while 70% of the ions are blocked. However, the attachment coefficient starts to rise after a few kV/cm thus limiting how much the $\vec{E}$-fields can be raised before too many $e^-$ are lost to absorption.

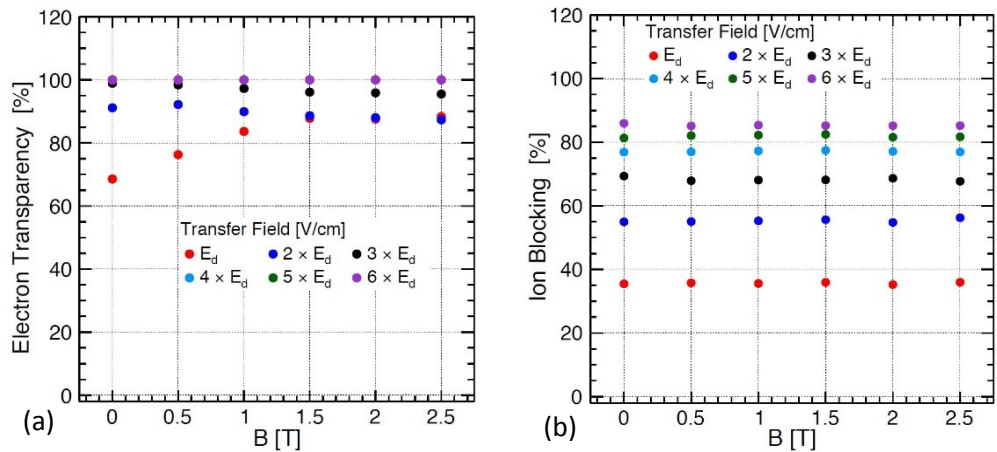

**Figure 3**: Simulation Results for Etched Mesh for (a) $e^-$ transparency & (b) ion blocking

There is even some ion blocking even at a field ratio of 1 because some of the ion drift lines will terminate at the mesh. These calculations show that a simple passive mesh placed before the gain structure could help run the amplification stage at lower voltages and improve its ion blocking capabilities. But blocking say twice as many ions, going from 35% to 70% from a field ratio of 1 to 3, in Fig. 3b, means that IBF is still the dominant source of SC. Earlier, our 2,000 gain and 1% IBF example gave a 20-to-1 ion ratio. This mesh scenario can lower it to 10-to-1, but it would only drop IBF from 95% to 90% from all SC.

The $\vec{B}$-field doesn't affect ions, as expected from their low drift velocity, but it does have an effect on the $e^-$. At lower $E$-field ratios the $\vec{B}$-field helps $e^-$ transparency by decreasing diffusion for $e^-$, as seen in Fig. 3a for a ratio of 1. However, following Fig. 3a to higher $\vec{B}$-fields, we see that all ratios pinch to a certain point. This is because in the limit of zero diffusion the $e^-$ travel straight and will saturate towards optical transparency.

## 3. Bi-Polar Static Wire Grid

An alternative to mesh structures is a parallel grid of wires with alternating voltages. Ions would be collected on negative wires, giving the possibility for full ion blocking, at appropriate arrangements. However, this can potentially lead to a large fraction of the $e^-$ to be collected on the positive wires. A potential solution is to continuously change the voltages on the wires between collisions based on the different $e^-$ and ion drift times. The bi-polar grid is "open" while $e^-$ are flowing through and "closed" when the ions are being collected. This was successfully done by STAR [1] to achieve their physics goals. But this active gating scheme isn't suitable for high-rate experiments since it creates long dead times. This will limit the capabilities of future high-rate collider experiments.

For a passive gate, where the potentials are not changed between collisions, without a $\vec{B}$-field most charges will be collected at the wires. Fig. 4a. shows this in a configuration of no field leaks where the orange lines are $e^-$ and grey lines are ions. Usually, heavy ion experiments want to minimize the $\vec{E} \times \vec{B}$ effect in order to minimize distortions, which happens globally. Since charges follow $\vec{E}$-field lines, hence $\hat{v} \parallel \hat{E}$, near the wires their velocities will no longer be parallel to the $\vec{B}$-field as they're attracted to the bi-polar wires. This allows us to exploit the $q(\vec{v} \times \vec{B})$ term from eq. 1 to help the $e^-$ pass through the bi-polar wires. This is done by creating a local kick on them while unaffecting the ions, which move too slowly to appreciably feel this magnetic force. Once they pass the wires, the $e^-$ will simply follow the $\vec{E}$ transfer lines to the gain stage. This effect is seen in Fig. 4b with only the $e^-$ lines shown. They pass through the bi-polar wires in the drift region at Z=0cm and are also seen to pinch between the wires.

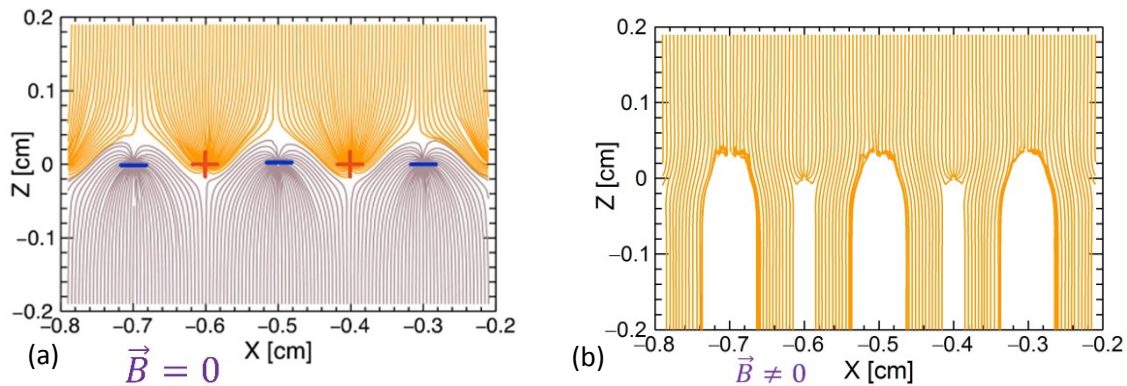

**Figure 4**: Bi-Polar wires (a) without and (b) with a $\vec{B}$-field

We've explored the parameter space over various factors such as: drift and transfer fields, wire diameters & separations between wires, whether or not wires of alternating voltages are in the same plane, and gas mixtures. The particular simulation configuration from Fig. 5 shows that the $e^-$ transparency can reduce to 90% from 100% (at $\pm V_{mesh} = 70[V]$), since not every $e^-$ would have a strong enough kick to make it past the bi-polar wires, but the ion blocking can jump by as much as 20%. This is a significant result, particularly when we mentioned earlier that 1 primary can produce 20 ions. With 90% of those produced ions being blocked we can have an almost 1-to-1 result of primaries-to-IBF.

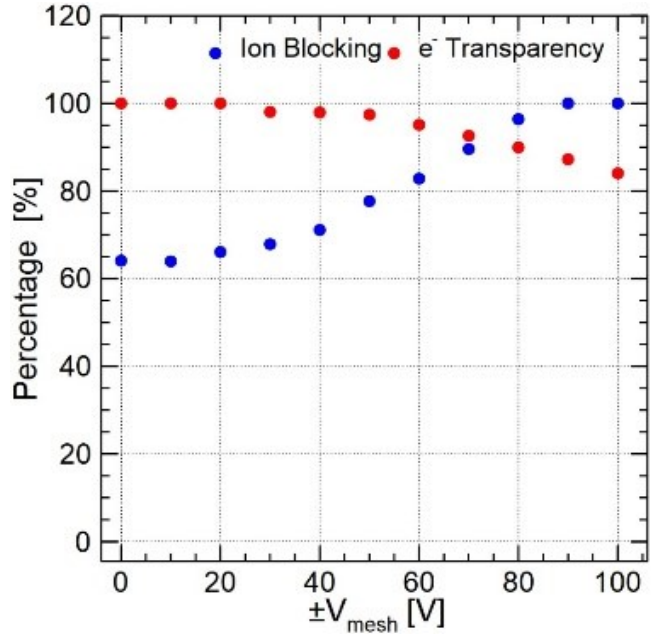

**Figure 5**: Simulation results in percentage {using Ne:CF$_4$ of 90:10, 300V/cm drift, 900V/cm transfer, 90μm wire diameter, 1.5mm wire pitch, & 60μm wire-plane split}

Furthermore, there are additional undesirable effects besides the primary $e^-$ loss. The $\hat{v} \times \hat{B}$ direction near the wire is also along the wire. And once the $e^-$ picks up a velocity component along the wire, it starts to move transversely to the wire and creates distortions along it. Fig. 6 (a, b) present a 3-D simulation of Fig. 4 (a, b), with the pitched & out-of-plane wire distortions emphasized by zooming in to the $e^-$ landing positions in Fig. 6c.

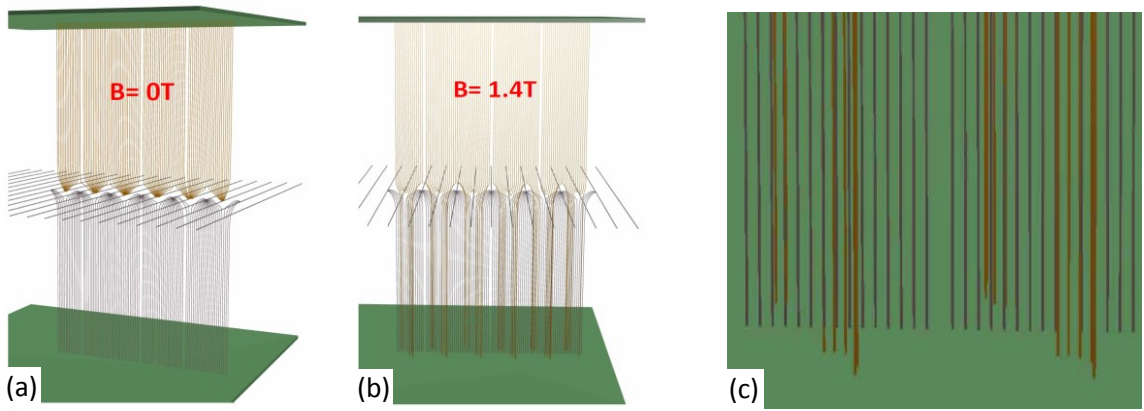

**Figure 6**: 3-D Simulation of charge transfer through Bi-Polar wires (a) without and (b) with a $\vec{B}$-field.

(c) zoomed-in orange $e^-$ lines landing "out of plane"

We now present the concept for correcting these wire distortions. It's seen from Fig. 4b that the $e^-$ displacements from their ideal trajectory are cyclical and that the cycle repeats with the same period as the bi-polar wires. Such cyclical shifts from ideal positions are known as DNL, and much effort has gone into minimizing DNL from the unique Zig-Zag pads. Our distinct concept is to design new pad shapes that are already specifically pre-distorted to compensate for this DNL from the new $e^-$ position caused by the "out-of-plane" shift due to the bi-polar wires.

### 4. Modified Pad Shape

Zig-Zag shapes with minimal DNL, as discussed in Ref. [6], are made by using the maximum "incursion" of neighboring pads and with a minimal tip-to-tip spacing, where it's less than the spot size of the gain-medium avalanche. The following is one approach to begin our investigations through simulations into the appropriate shape. We match the wire pitch to the pad pitch. We generate $e^-$ at positions that *should* intercept the gaps between Zig-Zag pads. We propagate them through all distortions, with emphasis on distortions caused by the bi-polar passive wires. And finally, we match the pad shapes and gaps to the $e^-$ landing spots as determined by these calculations.

This is a completely new approach since usually experiments would collect their data as-is and later, in the analysis stage, correct for any problems, such as distortions. But we are proposing to build pads that correct for distortion in the collection instead of the analysis stage. Fig. 7a shows a possible wires & pads configuration, as polarity of the wires can be flipped, or their positions relative to the pads and/or each other can be shifted, etc. More studies are currently ongoing to further optimize these configurations. Fig. 7 (b, c) show $e^-$ positions before and after they were propagated through the wire distortions and without any diffusion.

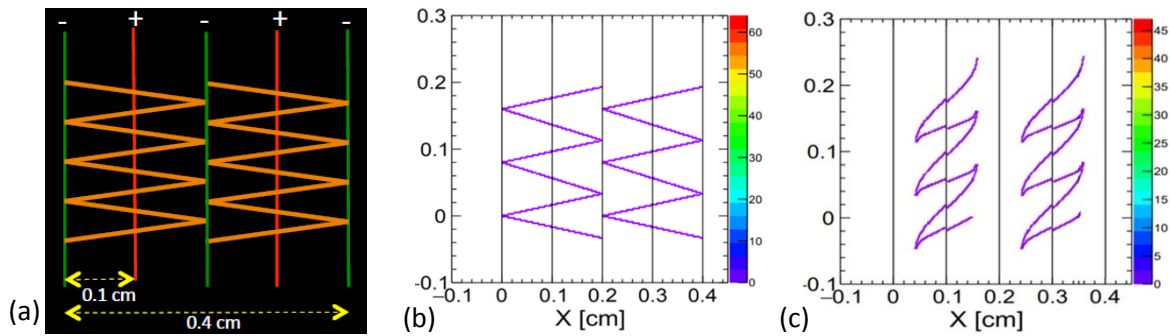

**Figure 7**: (a) possible wire-to-pad configuration. Simulations of (b) initial and (c) final $e^-$ positions propagated through the wires, without any diffusion, showing original and distorted pad edges

From the final $e^-$ locations in Fig. 7c, one proposed modified Zig-Zag pad would be the one in Fig. 8a. An ensemble of them to form a new detection read-out plane would resemble the cartoon in Fig. 8b. This array is noticeably different from our current Zig-Zag pads, which are shown in a photograph in Fig. 8c. The gold ENIG color is the pads and the black color is the gap between pads, although the simulations are shown for ideal zero-gap pads. The main distinction from Fig. 8b to Fig. 8c is the curved tips, resulting from the "out-of-plane" wire distortions, as opposed to the straight-ones.

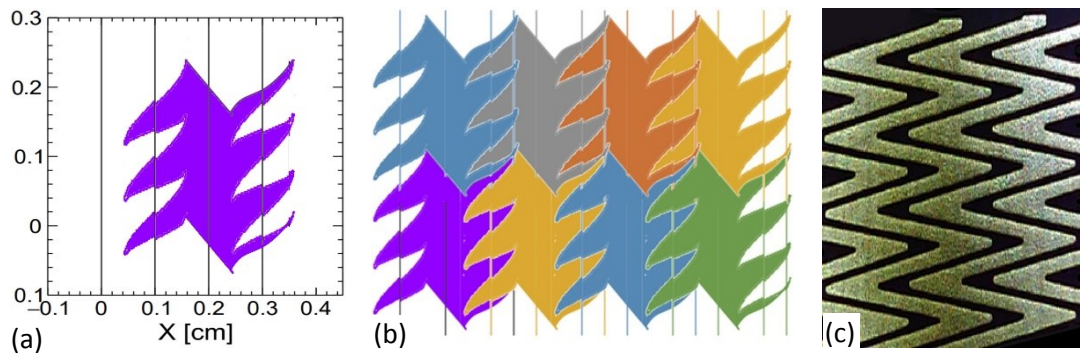

**Figure 8**: (a) possible shape for a single distorted pad, (b) an array of such distorted pads, (c) Photo of Zig-Zag pattern designed for the sPHENIX TPC

## 5. Conclusion

IBF is, by far, the main contributor to SC. Certain gain structures, such as a 4-GEM or 2-GEM-µMega are able to reduce IBF by creating a large transfer to drift $E$-field ratio. They also have other intrinsic properties, such as hole miss-alignment in Quad-GEMs, that help with IBF reduction. But such structures can cause fluctuations and deteriorate the detector performance. Passive structures, such as meshes, placed before the gain stage can greatly reduce IBF without hurting resolution, and our particular bi-polar wire grid simulations are showing reductions that will make IBF almost as low as the primary ion charges. The grid can distort $e^-$ position in a DNL manner, which might be accounted for by pre-distorted DNL pads. Further work will be done with more simulations of various parameters. The best configurations obtained by our simulation studies will be tested with a prototype detector.